\documentstyle[12pt, epsfig,aasms4]{article}

 1

\def\upm{{\stackrel{\raise.5ex\hbox{$m$}}{\lower.2ex\hbox{.}}}}
\def\upd{{\stackrel{\raise.5ex\hbox{$d$}}{\lower.2ex\hbox{.}}}}

\def\gs{\mathrel{\raise1.16pt\hbox{$>$}\kern-7.0pt
\lower3.06pt\hbox{{$\scriptstyle \sim$}}}}
\def\ls{\mathrel{\raise1.16pt\hbox{$<$}\kern-7.0pt
\lower3.06pt\hbox{{$\scriptstyle \sim$}}}}
\def\gtsima{$\; \buildrel > \over \sim \;$}
\def\ltsima{$\; \buildrel < \over \sim \;$}
\def\prosima{$\; \buildrel \propto \over \sim \;$}
\def\gsim{\lower.5ex\hbox{\gtsima}}
\def\lsim{\lower.5ex\hbox{\ltsima}}
\def\simgt{\lower.5ex\hbox{\gtsima}}
\def\simlt{\lower.5ex\hbox{\ltsima}}
\def\simpr{\lower.5ex\hbox{\prosima}}

\def\pp{\noindent\parshape 2 0truecm 17truecm 2truecm 15truecm}
\def\rf#1;#2;#3;#4 {\par\pp#1, #2, #3, #4. \par}

\def\pr{\ref@jnl{Phys.Rev}}

\def\href#1;#2 {{\bf #1} : {\em #2}}

\def\beq#1{\begin{equation}\label{#1}}
\def\eeq{\end{equation}}
\def\beqa#1{\begin{eqnarray}\label{#1}}
\def\eeqa{\end{eqnarray}}



\begin{document}
\thispagestyle{empty}
\title{  THE INSTABILITY STRIP FOR PRE--MAIN-SEQUENCE STARS}
 
\author {Marcella Marconi\altaffilmark{1,2} and 
Francesco Palla\altaffilmark{3} }

\altaffiltext{1}{Universit\`a degli studi di Firenze, Dipartimento di
Astronomia, L.go E. Fermi 5, 50125 Florence, Italy}
\altaffiltext{2}{Osservatorio Astronomico di Capodimonte, Via Moiariello
16, 80131 Napoli, Italy -
marcella@cerere.astrna.astro.it}
\altaffiltext{3}{Osservatorio Astrofisico di Arcetri, L.go E. Fermi 5, 
50125 Florence, Italy - palla@arcetri.astro.it}

\slugcomment{Astrophysical Journal {\it Letters}, in press}

\begin{abstract}

We investigate the pulsational properties of Pre--Main-Sequence (PMS) 
stars by means of linear and nonlinear calculations. The equilibrium models
were taken from models  evolved from the protostellar birthline to the ZAMS for masses in
the range 1 to 4 M$_\odot$. The nonlinear analysis allows us to define the
instability strip of PMS stars in the HR diagram.  These models are used to
constrain the internal structure of young stars and to test evolutionary
models. We compare our results with observations of the best case of a
pulsating young star, HR~5999, and we also identify possible candidates for
pulsational variability among known Herbig Ae/Be stars which are located
within or close to the instability strip boundaries.
\end{abstract}

\section{Introduction}

Young stars are characterized by a large degree of activity. Winds, jets
and outflows are manifestations of the interaction of the stars
with the circumstellar environment in which they are embedded. Similarly,
both T Tauri and Herbig Ae/Be stars show photometric and spectroscopic
variability on time scales of minutes to years, indicating that 
photospheric activity begins in the earliest
phases of stellar evolution, prior to the arrival on the Main Sequence (MS).
In many cases, strong and rapid line variability has been detected 
with periods
close to the star rotation period, interpreted as due to the rotational
modulation of spots on the stellar surface or of fast and slow streams
in the stellar wind (e.g. Gahm et al. 1995; Catala et al. 1989). 
On the other hand, the fact that young stars during their evolution to the
MS move across the instability region of post-MS stars
raises the possibility that at least part of the activity could also be
due to stellar pulsations (see Baade \& Stahl 1989; Kurtz \& Marang 1995).
However, the instability properties of young stars have not been studied
so far and it is not at all clear whether a Pre--Main-Sequence star
can indeed pulsate.

As recently discussed by Gautschy \& Saio (1996), 
stars over essentially the whole mass spectrum 
can become pulsationally unstable during various stages of their evolution. 
In particular, $\delta$ Scuti stars are intermediate-mass variables
of spectral type A to F with pulsation periods less than $0\upd 3$ and
light amplitudes ranging from thousandths of magnitude to some tenths
(Breger 1979; Breger \& Pamyatnykh 1998).
In spite of the traditional association
of $\delta$ Scuti pulsation with stars in their core hydrogen-burning phase 
or evolving towards the base of the giant branch and
burning hydrogen in a shell, 
some $\delta$ Scuti have been identified to be 
PMS objects (Breger 1972, Kurtz \& Marang 1995). 
Kurtz \& Marang (1995) have observed
the Herbig Ae star HR~5999 searching for pulsational photometric light
variations characteristic of $\delta$ Scuti stars. They have found 
a peak-to-peak
pulsation amplitude of about  $0\upm 013$ in Johnson V-band and a period 
of 4.99~hr, superimposed on a nonperiodic photometric variation of $0\upm 35$ 
probably due to  variations of dust obscuration in 
the circumstellar envelope and disk. 
There are other stars in which evidence for
pulsational variability has been advocated to explain the
observed rapid line profile variability (e.g. HD~163296 and
HD~250550, Baade \& Stahl 1989), but the case is not as convincing as that
provided by HR~5999. In any case, the discovery of pulsation in PMS stars
is extremely important since it provides a unique means for
constraining the internal structure of young stars and for testing
evolutionary models.

In this Letter, we explore the pulsational behavior of stellar
models distributed along the evolutionary tracks of PMS stars computed
by Palla \& Stahler (1993). We identify the boundaries of the
instability strip by using nonlinear models.
We can then use these models to constrain the fundamental stellar parameters
of HR~5999 and to select a sample of known PMS stars that are likely
candidates for showing pulsation variability in their photometric
light curves.

\section{Pre-main-sequence evolutionary models}\label{pms}

The calculation of the pulsational instability requires the specification
of the stellar parameters, luminosity and effective temperature, for 
a given mass.
We have used the PMS models for low- and intermediate-mass stars computed 
starting at the birthline determined by the
protostellar accretion phase (Palla \& Stahler 1990, 1993). The 
development of protostar
theory has altered the concept of PMS evolution particularly for stars
more massive than solar. In comparison with classical evolutionary tracks that
used arbitrary initial conditions, those of Palla \& Stahler (1993)
occupy a much reduced portion of the HR diagram as a consequence of the 
modest stellar radii attained by each star during the accretion period.
In addition, stars between 2 and 4 M$_\odot$ are not initially convective from
surface cooling, but begin their evolution in a thermally unrelaxed state
corresponding to luminosities and effective temperatures lower than those
of classical evolutionary tracks. 
This sequence of events may have direct effects on the study of
the instability properties of PMS stars and should be taken into account 
in the selection of the equilibrium models.

As an illustration of the differences in the internal structure of stars
at various stages of evolution, 
Fig.~1 shows the variation with radius of the interior density 
(normalized to the central value) for a 
3 M$_\odot$ star. On the birthline, at the beginning
of the PMS phase, the density profile is quite
flat and the star is thermally unrelaxed.
At the end of the relaxation phase, in the middle of the PMS track,
the star has already started its contraction to the ZAMS and is more
centrally condensed. The middle curve of Fig.~1 is for a model with 
luminosity
L$_\ast$=57~L$_\odot$ and effective temperature T$_{\rm eff}$=6900~K at 
an age of $7\times 10^5$~yr.
As we shall see below, such a model falls within the boundaries
of the instability strip.
Finally, the lower curve is for a star 
that has completed the main-sequence phase and is evolving toward
the red giant branch. The effective temperature is approximately the 
same as for the PMS star, but the luminosity is now 
L$_\ast$=135~L$_\odot$ due to 
the larger radius (8.3 R$_\odot$ instead of 5.3 R$_\odot$). These 
conditions correspond to those of a typical, evolved $\delta$~Scuti star. 
The star is much more centrally condensed because of the complete
exhaustion of hydrogen. However, beyond the inert core, the
density profile is quite similar to that of the PMS star. Thus, a mature
$\delta$~Scuti star differs from a young one mostly in the inner regions
and for its higher luminosity. These structural changes clearly
affect the estimate of the 
pulsation periods, as we shall see in the following sections.

\section{Pulsational models}\label{pulsa}

The theoretical framework for the investigation of the linear and nonlinear
pulsation characteristics has been described in several papers 
(e.g. Bono et al. 1997 and references therein) and is briefly 
summarized here.
We have computed several sequences of linear nonadiabatic models 
at fixed mass 
covering a wide range of luminosities and effective temperatures.
These models give information about periods and modal stability and
provide the static envelope structures for the nonlinear computations.
We adopt an optical
depth of the outermost zone of about 0.001 and fix the
inner boundary condition at a fraction of the stellar radius varying
between 10 and 20\%.
The zone closest to the Hydrogen Ionization Region (HIR) was
constrained to T$_{\rm HIR}=1.3\times 10^4$ K and 20 mass zones were
inserted between the HIR and the surface in order to ensure a good
spatial resolution of the outermost regions throughout the pulsation cycle.

Due to the key role played by spatial resolution for firmly estimating
both linear and nonlinear modal stability of higher modes, the envelopes
were discretized by adopting a detailed zoning in mass. The mass
ratio between consecutive zones ($\theta$) has been assumed equal to
$\theta$=1.1 for temperatures lower than $6 \times 10^5$ K and
to $\theta$=1.2 for higher temperatures.
In this way a typical envelope model, used for both linear and nonlinear
computations, is characterized
by roughly 20-30\% of the total stellar mass and by 150-250 zones.

To model the physical mechanisms which govern the pulsation phenomenon,
the $\kappa$ and $\gamma$ mechanisms in the hydrogen and helium ionization 
regions (see Cox 1980), 
we have adopted the OPAL radiative opacities provided by Rogers \& Iglesias
(1992) for temperatures higher than $10^4$~K and the molecular opacities 
of Alexander \& Ferguson (1994) for lower temperatures.
A solar chemical composition
(Y=0.28, Z=0.02) is assumed in all the computations.

\section{Linear analysis}

For each selected pulsation mode, the fundamental observables that 
can be derived from the linear nonadiabatic computations are periods and 
growth rates. 
In this work, we limited the study to the first three radial modes
of pulsation.
A detailed analysis of the sign of growth rates provides an estimate
of the location of the blue boundaries of the instability strips
for each mode, representing the maximum effective temperatures
allowing pulsation at a given luminosity level. 
No information is derived about the location of red boundaries
because our linear models are completely radiative while 
the quenching of pulsation at lower effective temperatures is 
due to the interaction between convective and dynamical motions
(e.g. Baker \& Kippenhan 1965). 

Although our goal is the definition of the theoretical instability strip,
we have specialized the linear analysis to the case of HR~5999.
In particular, we begin by reproducing its observed pulsation period
of 4.99 hr for a combination of mass, luminosity and effective
temperature consistent with both observations and evolutionary
tracks. To this purpose, we have computed a grid of linear nonadiabatic
models of stars with masses equal to 1.5, 2, 2.5, 3, 3.5, and 4 M$_{\odot}$,
luminosities in the range 10 to 140 L$_{\odot}$, and effective temperatures
4500$<$T$_{\rm eff}<$9500~K. The linear pulsation code assumes that the 
stellar models are in complete thermal and hydrostatic equilibrium.

Figure~2 shows the location
in the HR diagram of the lines of constant period ($P=4.99$ hr) for the
three pulsation modes and for different stellar masses.  At a given mass,
each line interesects the corresponding evolutionary track in one point that
determines the values of L$_\ast$ and T$_{\em eff}$ to be used in the
nonlinear models.  Also shown in Fig.~2 is the position of HR~5999 with the
observational uncertainties. The distance to the star based on Hipparcos
parallaxes is $d=210^{+50}_{-30}$~pc which gives a luminosity of
78$^{+37}_{-23}$~L$_\odot$ (van den Ancker et al. 1997). The spectral
classification of HR~5999 is A7III/IVe and estimates of the effective
temperature for this spectral type vary between $\sim$7600~K and $\sim$7000~K
(e.g. van den Ancker et al. 1997; Solano \& Fernley 1997).
As one can see from Fig.~2, only few equilibrium models with
the correct period fall within the HR~5999 box.  In particular, the model
with mass M$_\ast$=3 M$_\odot$ pulsates in the fundamental mode, that with
M$_\ast$=3.5 M$_\odot$ in the first overtone, and that with M$_\ast$=4
M$_\odot$ in the second overtone.  The selection of these models restricts
the choice of acceptable stellar masses for the detailed nonlinear analysis.

\section{Nonlinear analysis}

Given the static envelope structures produced by the linear computations,
we have performed the nonlinear stability analysis.
The equations governing both the
dynamical and convective structures were integrated in time until
the initial perturbations and the nonlinear fluctuations, which result from
the superposition of higher order modes, settled down (for more details see
Bono, Castellani \& Stellingwerf 1995).
Because of the very low growth rates (between $10^{-5}$ and $10^{-4}$), 
an extremely large period number (exceeding 10,000) must be calculated 
for each model
before reaching the stable limit cycle of pulsation (maximum amplitude).
The initial velocity profile was obtained
by perturbing the linear radial eigenfunctions with starting surface 
velocities varying in the range 0.1 to 10 km s$^{-1}$. 

As found in investigations of Post--Main-Sequence
$\delta$ Scuti star models (Milligan \& Carson 1992; Bono et al. 1997),  
the modal stability analysis
reveals the presence of stable nonlinear limit cycles in the fundamental, 
first and second overtone as the effective temperature increases.
Figure 3 shows the location of the theoretical instability 
strip boundaries in the HR diagram, together with the last static and 
the first pulsating
models found for different masses. The stellar parameters along the 
strip are listed in Table~1. 
In agreement with the distribution originally suggested by Breger \& 
Bregman (1975) and whith previous theoretical predictions 
(Bono et al. 1997), we find that second and first overtone pulsators 
are located at effective
temperatures higher than the fundamental ones.
All the models on the red edge are pulsating in the
fundamental mode, while those on the blue edge in either 
of the overtone modes.
The {\it red} edge varies between $\sim$6500~K and 
$\sim$7100~K, whereas the {\it blue} edge between $\sim$7100~$K$ and 
$\sim$7500~K. The width of the instability strip is nearly constant and 
equal to about 650~K. The time spent by each star within the strip
is also a constant fraction of the total PMS contraction time, typically
between 5 and 10\%, although in absolute terms it decreases from 
$\sim$10$^6$~yr for M$_\ast$=1.5 M$_\odot$ to 8$\times$10$^4$~yr 
for M$_\ast$=4 M$_\odot$. 

As shown in Fig.~3, the PMS instability strip terminates abruptly at the
blue edge of the M$_\ast$=4 M$_\odot$ evolutionary track. This results 
is the consequence of our assumption that
PMS stars begin their contraction phase at the birthline computed with
an accretion rate of $\dot M_{\rm acc}=10^{-5}$ M$_\odot$ yr$^{-1}$. 
It has been shown that such birthline reproduces quite well the 
the distribution of known PMS stars in the HR diagram (Palla \& Stahler 1990).
Obviously, the theoretical instability strip extends to luminosities 
higher than the upper limit found here for PMS stars, but it would then be
populated only by more evolved objects such as the classical $\delta$
Scuti stars.

\section{Discussion}\label{disc}

The models presented in this paper represent the first attempt to derive
the boundaries of the instability strip for young stars. A complete
analysis  of the predicted light curves and amplitudes and of the
sensitivity to the input physics will be given elsewhere (Marconi et al. 
1998, in preparation). However, we can use these initial results to set
constraints on the pulsational properties of known PMS objects.
The location of HR~5999 with respect to the instability strip is shown
in panel (b) of Fig.~3. We can see that in order to fall within the strip,
the effective temperature of HR~5999 should be lower than $\sim$7250~K if 
its mass is 3 M$_\odot$ or equal to $\sim$7100~K for M$_\ast$=4 M$_\odot$. 
However, if we accept the pulsation period of 4.99 hr
found by Kurtz \& Marang (1995), then the only acceptable models
would be (see Fig.~2) a 3 M$_\odot$ star pulsating in the fundamental mode,
a 3.5 M$_\odot$ in the first overtone, or a 4 M$_\odot$
star in the second overtone. On the other hand, these three models 
are located at the blue edge of 
the strip and the nonlinear stability analysis predicts that they should
all  pulsate in the second overtone mode. But in this higher mode only
the 4 M$_\odot$ model oscillates with the correct period of HR~5999,
ruling out the 3 M$_\odot$ and 3.5 M$_\odot$ stars as acceptable models.
Thus, we conclude that the
pulsation properties of HR~5999 suggest a mass of 4 M$_\odot$ with
second overtone pulsation. 

Despite the relatively short time spent by PMS stars inside the strip,
Fig.~3 shows that there is a number of known objects with the right
combination of luminosity and temperature to qualify as candidate pulsators.
These include five stars (V376~Cas, CQ~Tau, LkH$\alpha$~208, HD~142666 and
KK~Oph) of the
UX Orionis class, characterized by large photometric and polarimetric 
variations attributed to variable extinction by circumstellar dust
(Grinin \& Tambovtseva 1995; Natta et al. 1997). As in the case of HR~5999,
these stars should show periodic variations of smaller amplitudes in
their photometric behavior. Also, there are eight Herbig Ae/Be stars
of spectral type between A7 and F7 (HD~35929, V351~Ori, HD~142527, 
LkH$\alpha$~233, TY~CrA, RCr~A, TCr~A and HD~104237)
which are located inside or near the
instability strip within the uncertainties of the determination of 
L$_\ast$ and T$_{\rm eff}$ (Berrilli et al. 1992; van den Ancker et al. 1998). 
Searches for $\delta$ Scuti type variability in these stars are 
now under way.

\acknowledgments It is a pleasure to thank G. Bono and L. Terranegra for 
useful discussions on pulsation models and spectral properties of young stars.
This project was partly supported by CNR grant 97.00018.CT02 and by
ASI grant ARS 96-66 to the Osservatorio di Arcetri.

\newpage
 
%
%

\figcaption{The internal density profile (normalized to the central
value) of a 3 M$_\odot$ star in different evolutionary stages. The {\it upper
curve} is a protostar on the birthline computed with 
$\dot M_{\rm acc}=$10$^{-5}$~M$_\odot$~yr$^{-1}$ (Palla \& Stahler 1990). The
{\it middle curve} is for a pre-MS star with luminosity 
L$_\ast$=55 L$_\odot$, 
effective temperature T$_{\rm eff}$=6900 K, radius R$_\ast$=5.3 R$_\odot$ 
and an age of t$_{\rm PMS}=7\times 10^5$~yr.
Finally, the {\em lower curve} is for a post-MS star with the same effective
temperature, but L$_\ast$=135 L$_\odot$ and R$_\ast$=8.3 R$_\odot$. }

\figcaption{Location of the lines of constant period (equal to
4.99 hr) in the HR diagram. Three sets of lines are shown for the fundamental,
first and second overtone of stars of different masses.
The four curves of the fundamental mode are computed
for M$_\ast$=2.5, 3, 3.5 and 4 M$_\odot$. The three curves for the first
overtone refer to M$_\ast$=3, 3.5 and 4 M$_\odot$. The last two curves
for the second overtone are for M$_\ast$=3.5 and 4 M$_\odot$. 
The {\it filled dots} mark the intersection of the constant period lines 
with the corresponding
evolutionary tracks of Palla \& Stahler (1993), labelled by the appropriate
mass. The {\it long dashed curve} is the stellar birthline computed
with an accretion rate of $\dot M_{\em acc}=10^{-5}$ M$_\odot$~yr$^{-1}$.
Finally, the position of HR~5999 is indicated by the two asterisks, together
with the observational uncertainties in L$_\ast$ and T$_{\rm eff}$.  }

\figcaption{{\it Left panel}: location of the instability
strip of PMS stars in the HR diagram. {\it Right panel}: distribution
of known PMS stars within or near the boundaries of the instability strip
which can be considered candidates for pulsational instability. 
{\it triangles} show the position
of five UX Orionis-type stars (from Natta et al. 1997); {\it squares} are
eight Herbig Ae/Be stars (from Berrilli et al. 1992; van den Ancker 1998);
the {\it pentagon} is for HR~5999 and its associated uncertainty.  }


\clearpage
%
%
\begin{deluxetable}{lccccccccc}
\footnotesize
\tablecaption{Parameters of the instability strip \label{table}}
\tablewidth{0pt}\tablehead{
\multicolumn{1}{l}{} & \multicolumn{4}{c}{Red Edge} & \multicolumn{1}{l}{} &
\multicolumn{4}{c}{Blue Edge}\\
\cline{2-5}\cline{7-10}\\
\colhead{Mass}& \colhead{L$_\ast$} & \colhead{T$_{\rm eff}$} &\colhead{Mode} &
 \colhead{Period} &\colhead{} & \colhead{L$_\ast$} & \colhead{T$_{\rm eff}$} & \colhead{Mode} & \colhead{Period}\\
\colhead{(M$_\odot$)}& \colhead{(L$_\odot$)} & \colhead{(K)}& \colhead{} & \colhead{(d)} &
\colhead{} & \colhead{(L$_\odot$)} & \colhead{(K)} & \colhead{} & \colhead{(d)}}
\startdata
1.5 & 7   & 7090 & f  & 0.0611 & &     &       &  &  \nl
2.0 & 18  & 6870 & f  & 0.1195 & & 21  &  7500 & II & 0.0690 \nl
2.5 & 32  & 6640 & f  & 0.2155 & & 37  &  7350 & II & 0.0925 \nl
3.0 & 52  & 6500 & f  & 0.2601 & & 62  &  7260 & II & 0.1299 \nl
3.5 & 77  & 6550 & f  & 0.3195 & & 95  &  7200 & II & 0.1721 \nl
4.0 &     &      &    &        & & 133 &  7100 & I  & 0.2736\nl
\enddata
\end{deluxetable}

\clearpage
%
%

\clearpage

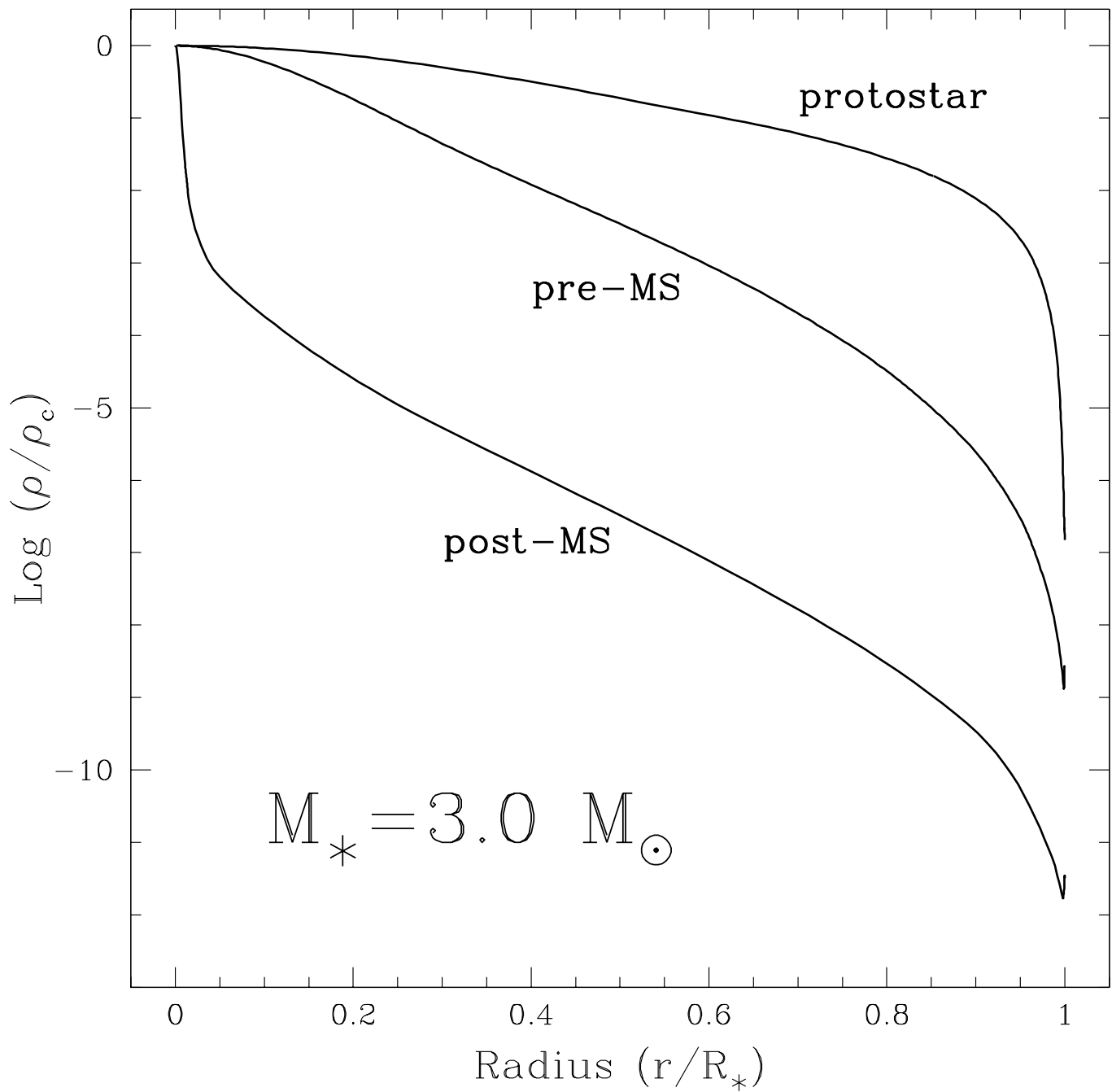
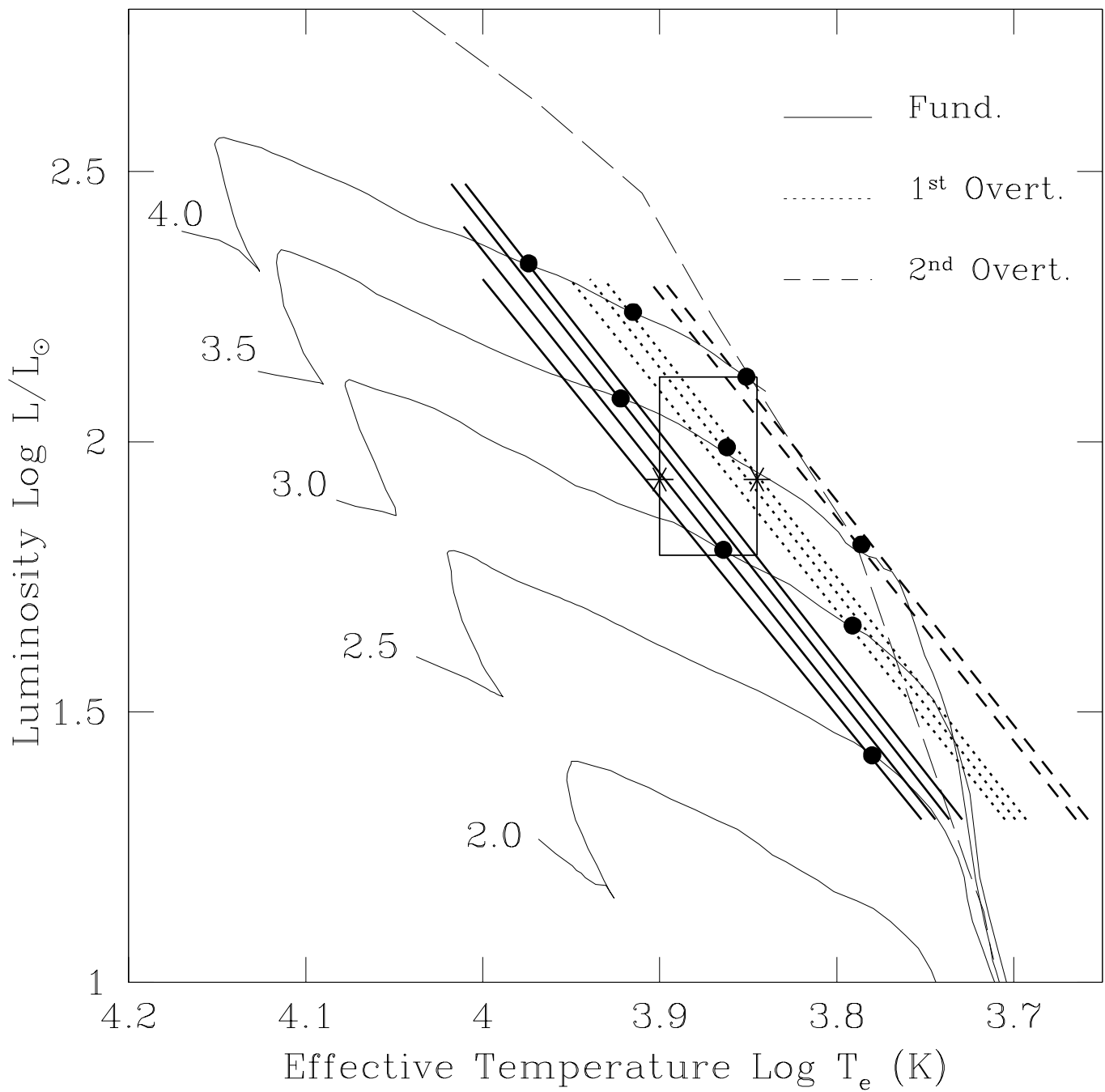
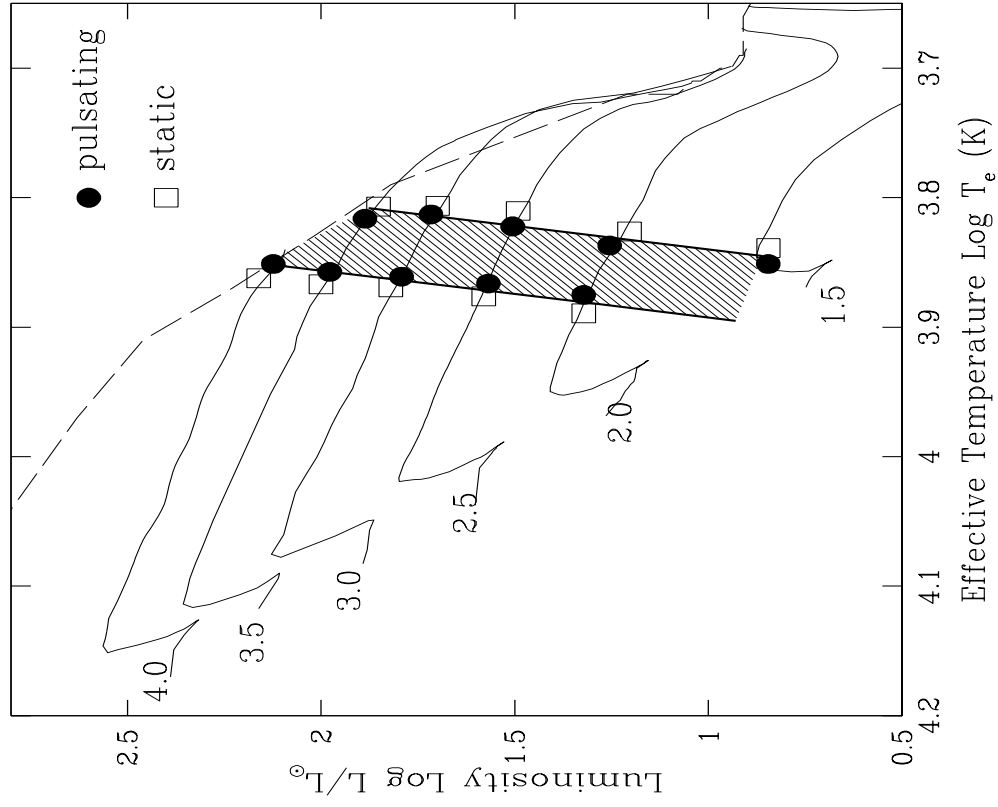
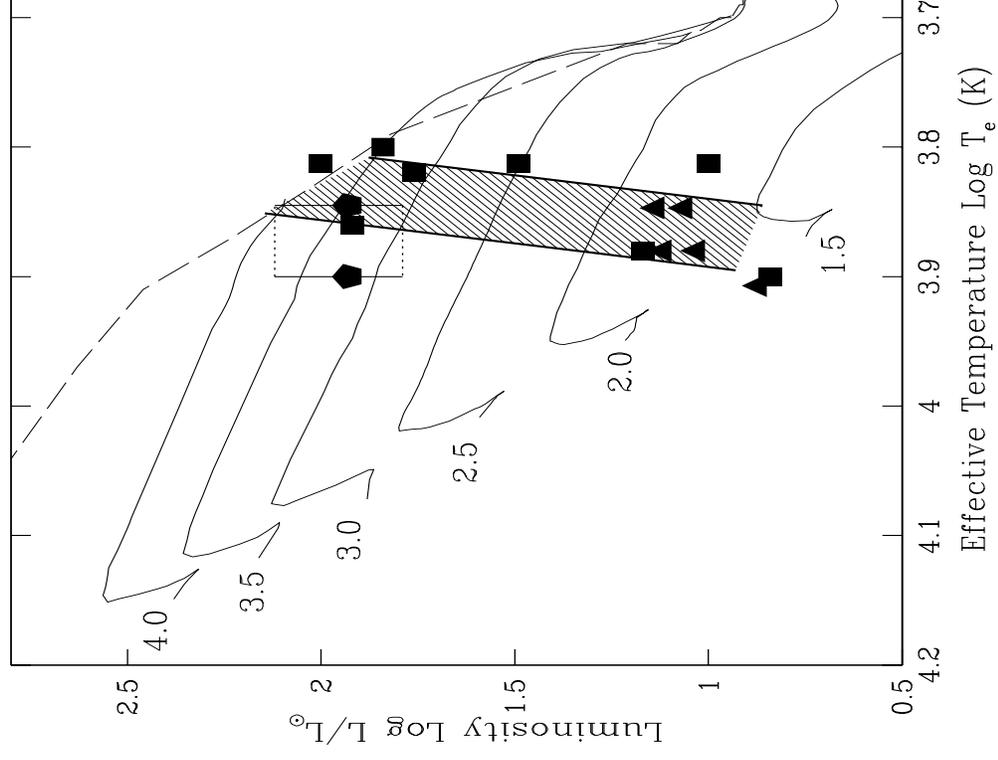
\end{document}